\documentclass[conference,10pt,twoside,twocolumn]{IEEEtran}

\usepackage{cite}
\usepackage[T1]{fontenc}
\usepackage{graphicx}
\usepackage{amssymb}
\usepackage{amsmath}
\usepackage{amsthm}

\usepackage{subfigure}
\usepackage{booktabs} 
\usepackage{multirow}
\usepackage{microtype}
\usepackage{fancyref}
\usepackage{xcolor}
\usepackage{balance}

\usepackage{amssymb}
\usepackage{amsfonts}
\usepackage{mathrsfs}
\usepackage{xspace}
\usepackage{bm}
\usepackage{upgreek}

\newcommand{\safemath}[2]{\newcommand{#1}{\ensuremath{#2}\xspace}}

\safemath{\bma}{\mathbf{a}}
\safemath{\bmb}{\mathbf{b}}
\safemath{\bmc}{\mathbf{c}}
\safemath{\bmd}{\mathbf{d}}
\safemath{\bme}{\mathbf{e}}
\safemath{\bmf}{\mathbf{f}}
\safemath{\bmg}{\mathbf{g}}
\safemath{\bmh}{\mathbf{h}}
\safemath{\bmi}{\mathbf{i}}
\safemath{\bmj}{\mathbf{j}}
\safemath{\bmk}{\mathbf{k}}
\safemath{\bml}{\mathbf{l}}
\safemath{\bmm}{\mathbf{m}}
\safemath{\bmn}{\mathbf{n}}
\safemath{\bmo}{\mathbf{o}}
\safemath{\bmp}{\mathbf{p}}
\safemath{\bmq}{\mathbf{q}}
\safemath{\bmr}{\mathbf{r}}
\safemath{\bms}{\mathbf{s}}
\safemath{\bmt}{\mathbf{t}}
\safemath{\bmu}{\mathbf{u}}
\safemath{\bmv}{\mathbf{v}}
\safemath{\bmw}{\mathbf{w}}
\safemath{\bmx}{\mathbf{x}}
\safemath{\bmy}{\mathbf{y}}
\safemath{\bmz}{\mathbf{z}}
\safemath{\bmzero}{\mathbf{0}}
\safemath{\bmone}{\mathbf{1}}

\bmdefine{\biad}{a}
\bmdefine{\bibd}{b}
\bmdefine{\bicd}{c}
\bmdefine{\bidd}{d}
\bmdefine{\bied}{e}
\bmdefine{\bifd}{f}
\bmdefine{\bigd}{g}
\bmdefine{\bihd}{h}
\bmdefine{\biid}{i}
\bmdefine{\bijd}{j}
\bmdefine{\bikd}{k}
\bmdefine{\bild}{l}
\bmdefine{\bimd}{m}
\bmdefine{\bind}{n}
\bmdefine{\biod}{o}
\bmdefine{\bipd}{p}
\bmdefine{\biqd}{q}
\bmdefine{\bird}{r}
\bmdefine{\bisd}{s}
\bmdefine{\bitd}{t}
\bmdefine{\biud}{u}
\bmdefine{\bivd}{v}
\bmdefine{\biwd}{w}
\bmdefine{\bixd}{x}
\bmdefine{\biyd}{y}
\bmdefine{\bizd}{z}

\bmdefine{\bixid}{\xi}
\bmdefine{\bilambdad}{\lambda}
\bmdefine{\bimud}{\mu}
\bmdefine{\bithetad}{\theta}
\bmdefine{\biphid}{\phi}
\bmdefine{\bideltad}{\delta}

\safemath{\bmia}{\biad}
\safemath{\bmib}{\bibd}
\safemath{\bmic}{\bicd}
\safemath{\bmid}{\bidd}
\safemath{\bmie}{\bied}
\safemath{\bmif}{\bifd}
\safemath{\bmig}{\bigd}
\safemath{\bmih}{\bihd}
\safemath{\bmii}{\biid}
\safemath{\bmij}{\bijd}
\safemath{\bmik}{\bikd}
\safemath{\bmil}{\bild}
\safemath{\bmim}{\bimd}
\safemath{\bmin}{\bind}
\safemath{\bmio}{\biod}
\safemath{\bmip}{\bipd}
\safemath{\bmiq}{\biqd}
\safemath{\bmir}{\bird}
\safemath{\bmis}{\bisd}
\safemath{\bmit}{\bitd}
\safemath{\bmiu}{\biud}
\safemath{\bmiv}{\bivd}
\safemath{\bmiw}{\biwd}
\safemath{\bmix}{\bixd}
\safemath{\bmiy}{\biyd}
\safemath{\bmiz}{\bizd}

\safemath{\bmxi}{\bixid}
\safemath{\bmlambda}{\bilambdad}
\safemath{\bmmu}{\bimud}
\safemath{\bmtheta}{\bithetad}
\safemath{\bmphi}{\biphid}
\safemath{\bmdelta}{\bideltad}

\safemath{\bA}{\mathbf{A}}
\safemath{\bB}{\mathbf{B}}
\safemath{\bC}{\mathbf{C}}
\safemath{\bD}{\mathbf{D}}
\safemath{\bE}{\mathbf{E}}
\safemath{\bF}{\mathbf{F}}
\safemath{\bG}{\mathbf{G}}
\safemath{\bH}{\mathbf{H}}
\safemath{\bI}{\mathbf{I}}
\safemath{\bJ}{\mathbf{J}}
\safemath{\bK}{\mathbf{K}}
\safemath{\bL}{\mathbf{L}}
\safemath{\bM}{\mathbf{M}}
\safemath{\bN}{\mathbf{N}}
\safemath{\bO}{\mathbf{O}}
\safemath{\bP}{\mathbf{P}}
\safemath{\bQ}{\mathbf{Q}}
\safemath{\bR}{\mathbf{R}}
\safemath{\bS}{\mathbf{S}}
\safemath{\bT}{\mathbf{T}}
\safemath{\bU}{\mathbf{U}}
\safemath{\bV}{\mathbf{V}}
\safemath{\bW}{\mathbf{W}}
\safemath{\bX}{\mathbf{X}}
\safemath{\bY}{\mathbf{Y}}
\safemath{\bZ}{\mathbf{Z}}

\safemath{\bZero}{\mathbf{0}}
\safemath{\bOne}{\mathbf{1}}
\safemath{\bDelta}{\mathbf{\Delta}}
\safemath{\bLambda}{\mathbf{\UpLambda}}
\safemath{\bPhi}{\mathbf{\Upphi}}
\safemath{\bSigma}{\mathbf{\Upsigma}}
\safemath{\bOmega}{\mathbf{\Upomega}}
\safemath{\bTheta}{\mathbf{\Uptheta}}

\bmdefine{\biAd}{A}
\bmdefine{\biBd}{B}
\bmdefine{\biCd}{C}
\bmdefine{\biDd}{D}
\bmdefine{\biEd}{E}
\bmdefine{\biFd}{F}
\bmdefine{\biGd}{G}
\bmdefine{\biHd}{H}
\bmdefine{\biId}{I}
\bmdefine{\biJd}{J}
\bmdefine{\biKd}{K}
\bmdefine{\biLd}{L}
\bmdefine{\biMd}{M}
\bmdefine{\biOd}{N}
\bmdefine{\biPd}{O}
\bmdefine{\biQd}{P}
\bmdefine{\biRd}{R}
\bmdefine{\biSd}{S}
\bmdefine{\biTd}{T}
\bmdefine{\biUd}{U}
\bmdefine{\biVd}{V}
\bmdefine{\biWd}{W}
\bmdefine{\biXd}{X}
\bmdefine{\biYd}{Y}
\bmdefine{\biZd}{Z}

\bmdefine{\biDelta}{\Delta}
\bmdefine{\biLambda}{\Lambda}
\bmdefine{\biPhi}{\Phi}
\bmdefine{\biSigma}{\Sigma}
\bmdefine{\biOmega}{\Omega}
\bmdefine{\biTheta}{\Theta}

\safemath{\bimA}{\biAd}
\safemath{\bimB}{\biBd}
\safemath{\bimC}{\biCd}
\safemath{\bimD}{\biDd}
\safemath{\bimE}{\biEd}
\safemath{\bimF}{\biFd}
\safemath{\bimG}{\biGd}
\safemath{\bimH}{\biHd}
\safemath{\bimI}{\biId}
\safemath{\bimJ}{\biJd}
\safemath{\bimK}{\biKd}
\safemath{\bimL}{\biLd}
\safemath{\bimM}{\biMd}
\safemath{\bimN}{\biNd}
\safemath{\bimO}{\biOd}
\safemath{\bimP}{\biPd}
\safemath{\bimQ}{\biQd}
\safemath{\bimR}{\biRd}
\safemath{\bimS}{\biSd}
\safemath{\bimT}{\biTd}
\safemath{\bimU}{\biUd}
\safemath{\bimV}{\biVd}
\safemath{\bimW}{\biWd}
\safemath{\bimX}{\biXd}
\safemath{\bimY}{\biYd}
\safemath{\bimZ}{\biZd}

\safemath{\bimDelta}{\biDelta}
\safemath{\bimLambda}{\biLambda}
\safemath{\bimPhi}{\biPhi}
\safemath{\bimSigma}{\biSigma}
\safemath{\bimOmega}{\biOmega}
\safemath{\bimTheta}{\biTheta}

\safemath{\setA}{\mathcal{A}}
\safemath{\setB}{\mathcal{B}}
\safemath{\setC}{\mathcal{C}}
\safemath{\setD}{\mathcal{D}}
\safemath{\setE}{\mathcal{E}}
\safemath{\setF}{\mathcal{F}}
\safemath{\setG}{\mathcal{G}}
\safemath{\setH}{\mathcal{H}}
\safemath{\setI}{\mathcal{I}}
\safemath{\setJ}{\mathcal{J}}
\safemath{\setK}{\mathcal{K}}
\safemath{\setL}{\mathcal{L}}
\safemath{\setM}{\mathcal{M}}
\safemath{\setN}{\mathcal{N}}
\safemath{\setO}{\mathcal{O}}
\safemath{\setP}{\mathcal{P}}
\safemath{\setQ}{\mathcal{Q}}
\safemath{\setR}{\mathcal{R}}
\safemath{\setS}{\mathcal{S}}
\safemath{\setT}{\mathcal{T}}
\safemath{\setU}{\mathcal{U}}
\safemath{\setV}{\mathcal{V}}
\safemath{\setW}{\mathcal{W}}
\safemath{\setX}{\mathcal{X}}
\safemath{\setY}{\mathcal{Y}}
\safemath{\setZ}{\mathcal{Z}}
\safemath{\emptySet}{\varnothing}

\safemath{\colA}{\mathscr{A}}
\safemath{\colB}{\mathscr{B}}
\safemath{\colC}{\mathscr{C}}
\safemath{\colD}{\mathscr{D}}
\safemath{\colE}{\mathscr{E}}
\safemath{\colF}{\mathscr{F}}
\safemath{\colG}{\mathscr{G}}
\safemath{\colH}{\mathscr{H}}
\safemath{\colI}{\mathscr{I}}
\safemath{\colJ}{\mathscr{J}}
\safemath{\colK}{\mathscr{K}}
\safemath{\colL}{\mathscr{L}}
\safemath{\colM}{\mathscr{M}}
\safemath{\colN}{\mathscr{N}}
\safemath{\colO}{\mathscr{O}}
\safemath{\colP}{\mathscr{P}}
\safemath{\colQ}{\mathscr{Q}}
\safemath{\colR}{\mathscr{R}}
\safemath{\colS}{\mathscr{S}}
\safemath{\colT}{\mathscr{T}}
\safemath{\colU}{\mathscr{U}}
\safemath{\colV}{\mathscr{V}}
\safemath{\colW}{\mathscr{W}}
\safemath{\colX}{\mathscr{X}}
\safemath{\colY}{\mathscr{Y}}
\safemath{\colZ}{\mathscr{Z}}

\safemath{\opA}{\mathbb{A}}
\safemath{\opB}{\mathbb{B}}
\safemath{\opC}{\mathbb{C}}
\safemath{\opD}{\mathbb{D}}
\safemath{\opE}{\mathbb{E}}
\safemath{\opF}{\mathbb{F}}
\safemath{\opG}{\mathbb{G}}
\safemath{\opH}{\mathbb{H}}
\safemath{\opI}{\mathbb{I}}
\safemath{\opJ}{\mathbb{J}}
\safemath{\opK}{\mathbb{K}}
\safemath{\opL}{\mathbb{L}}
\safemath{\opM}{\mathbb{M}}
\safemath{\opN}{\mathbb{N}}
\safemath{\opO}{\mathbb{O}}
\safemath{\opP}{\mathbb{P}}
\safemath{\opQ}{\mathbb{Q}}
\safemath{\opR}{\mathbb{R}}
\safemath{\opS}{\mathbb{S}}
\safemath{\opT}{\mathbb{T}}
\safemath{\opU}{\mathbb{U}}
\safemath{\opV}{\mathbb{V}}
\safemath{\opW}{\mathbb{W}}
\safemath{\opX}{\mathbb{X}}
\safemath{\opY}{\mathbb{Y}}
\safemath{\opZ}{\mathbb{Z}}
\safemath{\opZero}{\mathbb{O}}
\safemath{\identityop}{\opI}

\safemath{\veca}{\bma}
\safemath{\vecb}{\bmb}
\safemath{\vecc}{\bmc}
\safemath{\vecd}{\bmd}
\safemath{\vece}{\bme}
\safemath{\vecf}{\bmf}
\safemath{\vecg}{\bmg}
\safemath{\vech}{\bmh}
\safemath{\veci}{\bmi}
\safemath{\vecj}{\bmj}
\safemath{\veck}{\bmk}
\safemath{\vecl}{\bml}
\safemath{\vecm}{\bmm}
\safemath{\vecn}{\bmn}
\safemath{\veco}{\bmo}
\safemath{\vecp}{\bmp}
\safemath{\vecq}{\bmq}
\safemath{\vecr}{\bmr}
\safemath{\vecs}{\bms}
\safemath{\vect}{\bmt}
\safemath{\vecu}{\bmu}
\safemath{\vecv}{\bmv}
\safemath{\vecw}{\bmw}
\safemath{\vecx}{\bmx}
\safemath{\vecy}{\bmy}
\safemath{\vecz}{\bmz}

\safemath{\veczero}{\bmzero}
\safemath{\vecone}{\bmone}
\safemath{\vecxi}{\bmxi}
\safemath{\veclambda}{\bmlambda}
\safemath{\vecmu}{\bmmu}
\safemath{\vectheta}{\bmtheta}
\safemath{\vecphi}{\bmphi}
\safemath{\vecdelta}{\bmdelta}

\safemath{\matA}{\bA}
\safemath{\matB}{\bB}
\safemath{\matC}{\bC}
\safemath{\matD}{\bD}
\safemath{\matE}{\bE}
\safemath{\matF}{\bF}
\safemath{\matG}{\bG}
\safemath{\matH}{\bH}
\safemath{\matI}{\bI}
\safemath{\matJ}{\bJ}
\safemath{\matK}{\bK}
\safemath{\matL}{\bL}
\safemath{\matM}{\bM}
\safemath{\matN}{\bN}
\safemath{\matO}{\bO}
\safemath{\matP}{\bP}
\safemath{\matQ}{\bQ}
\safemath{\matR}{\bR}
\safemath{\matS}{\bS}
\safemath{\matT}{\bT}
\safemath{\matU}{\bU}
\safemath{\matV}{\bV}
\safemath{\matW}{\bW}
\safemath{\matX}{\bX}
\safemath{\matY}{\bY}
\safemath{\matZ}{\bZ}
\safemath{\matzero}{\bmzero}

\safemath{\matDelta}{\bDelta}
\safemath{\matLambda}{\bLambda}
\safemath{\matPhi}{\bPhi}
\safemath{\matSigma}{\bSigma}
\safemath{\matOmega}{\bOmega}
\safemath{\matTheta}{\bTheta}

\safemath{\matidentity}{\matI}
\safemath{\matone}{\matO}

\safemath{\rnda}{A}
\safemath{\rndb}{B}
\safemath{\rndc}{C}
\safemath{\rndd}{D}
\safemath{\rnde}{E}
\safemath{\rndf}{F}
\safemath{\rndg}{G}
\safemath{\rndh}{H}
\safemath{\rndi}{I}
\safemath{\rndj}{J}
\safemath{\rndk}{K}
\safemath{\rndl}{L}
\safemath{\rndm}{M}
\safemath{\rndn}{N}
\safemath{\rndo}{O}
\safemath{\rndp}{P}
\safemath{\rndq}{Q}
\safemath{\rndr}{R}
\safemath{\rnds}{S}
\safemath{\rndt}{T}
\safemath{\rndu}{U}
\safemath{\rndv}{V}
\safemath{\rndw}{W}
\safemath{\rndx}{X}
\safemath{\rndy}{Y}
\safemath{\rndz}{Z}

\safemath{\rveca}{\bimA}
\safemath{\rvecb}{\bimB}
\safemath{\rvecc}{\bimC}
\safemath{\rvecd}{\bimD}
\safemath{\rvece}{\bimE}
\safemath{\rvecf}{\bimF}
\safemath{\rvecg}{\bimG}
\safemath{\rvech}{\bimH}
\safemath{\rveci}{\bimI}
\safemath{\rvecj}{\bimJ}
\safemath{\rveck}{\bimK}
\safemath{\rvecl}{\bimL}
\safemath{\rvecm}{\bimM}
\safemath{\rvecn}{\bimN}
\safemath{\rveco}{\bomO}
\safemath{\rvecp}{\bimP}
\safemath{\rvecq}{\bimQ}
\safemath{\rvecr}{\bimR}
\safemath{\rvecs}{\bimS}
\safemath{\rvect}{\bimT}
\safemath{\rvecu}{\bimU}
\safemath{\rvecv}{\bimV}
\safemath{\rvecw}{\bimW}
\safemath{\rvecx}{\bimX}
\safemath{\rvecy}{\bimY}
\safemath{\rvecz}{\bimZ}

\safemath{\rvecxi}{\bmxi}
\safemath{\rveclambda}{\bmlambda}
\safemath{\rvecmu}{\bmmu}
\safemath{\rvectheta}{\bmtheta}
\safemath{\rvecphi}{\bmphi}

\safemath{\rmatA}{\bimA}
\safemath{\rmatB}{\bimB}
\safemath{\rmatC}{\bimC}
\safemath{\rmatD}{\bimD}
\safemath{\rmatE}{\bimE}
\safemath{\rmatF}{\bimF}
\safemath{\rmatG}{\bimG}
\safemath{\rmatH}{\bimH}
\safemath{\rmatI}{\bimI}
\safemath{\rmatJ}{\bimJ}
\safemath{\rmatK}{\bimK}
\safemath{\rmatL}{\bimL}
\safemath{\rmatM}{\bimM}
\safemath{\rmatN}{\bimN}
\safemath{\rmatO}{\bimO}
\safemath{\rmatP}{\bimP}
\safemath{\rmatQ}{\bimQ}
\safemath{\rmatR}{\bimR}
\safemath{\rmatS}{\bimS}
\safemath{\rmatT}{\bimT}
\safemath{\rmatU}{\bimU}
\safemath{\rmatV}{\bimV}
\safemath{\rmatW}{\bimW}
\safemath{\rmatX}{\bimX}
\safemath{\rmatY}{\bimY}
\safemath{\rmatZ}{\bimZ}

\safemath{\rmatDelta}{\bimDelta}
\safemath{\rmatLambda}{\bimLambda}
\safemath{\rmatPhi}{\bimPhi}
\safemath{\rmatSigma}{\bimSigma}
\safemath{\rmatOmega}{\bimOmega}
\safemath{\rmatTheta}{\bimTheta}

\usepackage{amssymb}
\usepackage{amsfonts}
\usepackage{mathrsfs}
\usepackage{xspace}
\usepackage{bm}
\usepackage{fancyref}
\usepackage{textcomp}

\usepackage{multirow}
\usepackage{stmaryrd}

\newenvironment{textbmatrix}{	\setlength{\arraycolsep}{2.5pt}%
								\big[\begin{matrix}}{\end{matrix}\big]%
								\raisebox{0.08ex}{\vphantom{M}}}

\def\be{\begin{equation}}
\def\ee{\end{equation}}
\def\een{\nonumber \end{equation}}
\def\mat{\begin{bmatrix}}
\def\emat{\end{bmatrix}}
\def\btm{\begin{textbmatrix}}
\def\etm{\end{textbmatrix}}

\def\ba#1\ea{\begin{align}#1\end{align}}
\def\bas#1\eas{\begin{align*}#1\end{align*}}
\def\bs#1\es{\begin{split}#1\end{split}}
\def\bg#1\eg{\begin{gather}#1\end{gather}}
\def\bml#1\eml{\begin{multline}#1\end{multline}}
\def\bi#1\ei{\begin{itemize}#1\end{itemize}}

\safemath{\dirac}{\delta}					% Dirac delta
\safemath{\krond}{\dirac}					% Kronecker delta
\safemath{\upto}{\uparrow}
\safemath{\downto}{\downarrow}
\safemath{\iu}{j}							% imaginary unit
\safemath{\ev}{\lambda}						% eigenvalue
\safemath{\hilseqspace}{l^{2}}				% Hilbert sequence space
\newcommand{\banachfunspace}[1]{\setL^{#1}}	% Banach function space
\safemath{\hilfunspace}{\banachfunspace{2}}	% Hilbert function space
\safemath{\SNR}{\textit{SNR}} 				% signal to noise ratio
\safemath{\PAR}{\textit{PAR}} 				% signal to noise ratio
\safemath{\No}{N_0}							% noise spectral density
\safemath{\Es}{E_s}							% energy per symbol
\safemath{\Eb}{E_b}							% energy per bit
\safemath{\EbNo}{\frac{\Eb}{\No}}
\safemath{\EsNo}{\frac{\Es}{\No}}

\DeclareMathOperator{\CHop}{\ensuremath{\opH}} % channel operator
\safemath{\tvir}{\rndh_{\CHop}}				% time-varying impulse response
\safemath{\tvtf}{\rndl_{\CHop}}				% 	-''- transfer function
\safemath{\spf}{\rnds_{\CHop}}				% spreading function
\safemath{\bff}{H_{\CHop}}					% bi-freuqency function

\safemath{\ircf}{r_{h}}						% impulse response correlation fn.
\safemath{\tftvcf}{r_{s}}					% scattering function
\safemath{\tfcf}{r_{l}}						% time-frequency correlation fn.
\safemath{\bfcf}{r_{H}}						% bi-frequency correlation fn.

\safemath{\tcorr}{c_h}						% time-correlation function
\safemath{\scf}{c_{s}}						% spreading function
\safemath{\tfcorr}{c_{l}}					% transfer-function correlation
\safemath{\fcorr}{c_{H}}						% frequency-correlation function

\safemath{\mi}{I}							% mutual information
\safemath{\capacity}{C}						% capacity

\safemath{\normal}{\mathcal{N}}			% normal distribution
\safemath{\jpg}{\mathcal{CN}}			% jointly proper Gaussian
\safemath{\mchain}{\leftrightarrow}		% Markov chain
\safemath{\dB}{\,\mathrm{dB}}
\safemath{\dBm}{\,\mathrm{dBm}}
\safemath{\Hz}{\,\mathrm{Hz}}
\safemath{\kHz}{\,\mathrm{kHz}}
\safemath{\MHz}{\,\mathrm{MHz}}
\safemath{\GHz}{\,\mathrm{GHz}}
\safemath{\s}{\,\mathrm{s}}
\safemath{\ms}{\,\mathrm{ms}}
\safemath{\mus}{\,\mathrm{\text{\textmu}s}}
\safemath{\ns}{\,\mathrm{ns}}
\safemath{\ps}{\,\mathrm{ps}}
\safemath{\meter}{\,\mathrm{m}}
\safemath{\mm}{\,\mathrm{mm}}
\safemath{\cm}{\,\mathrm{cm}}
\safemath{\m}{\,\mathrm{m}}
\safemath{\W}{\,\mathrm{W}}
\safemath{\mW}{\, \mathrm{mW}}
\safemath{\J}{\,\mathrm{J}}
\safemath{\K}{\,\mathrm{K}}
\safemath{\bit}{\,\mathrm{bit}}
\safemath{\nat}{\,\mathrm{nat}}

\safemath{\define}{\triangleq}			% definition

\safemath{\equivalent}{\sim}
\safemath{\distas}{\sim}					% distributed according to
\safemath{\sdiff}{\Delta}				% symmetric set difference

\safemath{\reals}{\mathbb{R}}
\safemath{\positivereals}{\reals_{+}}
\safemath{\integers}{\mathbb{Z}}
\safemath{\posint}{\integers_{+}}
\safemath{\naturals}{\mathbb{N}}
\safemath{\posnaturals}{\naturals_{+}}
\safemath{\complexset}{\mathbb{C}}
\safemath{\rationals}{\mathbb{Q}}

\newcommand*{\fancyrefapplabelprefix}{app}		% Appendix
\newcommand*{\fancyrefthmlabelprefix}{thm}		% Theorem
\newcommand*{\fancyreflemlabelprefix}{lem}		% Lemma
\newcommand*{\fancyrefcorlabelprefix}{cor}		% Corollary
\newcommand*{\fancyrefdeflabelprefix}{def}		% Definition
\newcommand*{\fancyrefproplabelprefix}{prop}		% Proposition
\newcommand*{\fancyrefexmpllabelprefix}{exmpl}
\newcommand*{\fancyrefalglabelprefix}{alg}		% Algorithm
\newcommand*{\fancyreftbllabelprefix}{tbl}		% Algorithm

\frefformat{vario}{\fancyrefseclabelprefix}{Section~#1}
\frefformat{vario}{\fancyrefthmlabelprefix}{Theorem~#1}
\frefformat{vario}{\fancyreftbllabelprefix}{Table~#1}
\frefformat{vario}{\fancyreflemlabelprefix}{Lemma~#1}
\frefformat{vario}{\fancyrefcorlabelprefix}{Corollary~#1}
\frefformat{vario}{\fancyrefdeflabelprefix}{Definition~#1}
\frefformat{vario}{\fancyreffiglabelprefix}{Fig.~#1}
\frefformat{vario}{\fancyrefapplabelprefix}{Appendix~#1}
\frefformat{vario}{\fancyrefeqlabelprefix}{(#1)}
\frefformat{vario}{\fancyrefproplabelprefix}{Proposition~#1}
\frefformat{vario}{\fancyrefexmpllabelprefix}{Example~#1}
\frefformat{vario}{\fancyrefalglabelprefix}{Algorithm~#1}

\safemath{\dictab}{[\,\dicta\,\,\dictb\,]}

\safemath{\ysig}{\bmy}
\safemath{\ysighat}{\hat{\ysig}}
\safemath{\ysigdim}{M}
\safemath{\xsig}{\bmx}
\safemath{\xsigdim}{N}
\safemath{\nx}{n_x}
\safemath{\zsig}{\bmz}
\safemath{\zsigdim}{\ysigdim}
\safemath{\rsig}{\bmr}
\safemath{\Adict}{\bA}
\safemath{\Adicttilde}{\widetilde{\Adict}}
\safemath{\Adictdim}{\outputdim\times\xsigdim}
\safemath{\avec}{\bma}
\safemath{\avectilde}{\tilde{\avec}}
\safemath{\Bdict}{\bB}
\safemath{\Bdicttilde}{\widetilde{\Bdict}}
\safemath{\Cdict}{\bC}
\safemath{\cvec}{\bmc}
\safemath{\Ddict}{\bD}
\safemath{\Ddictdim}{\ysigdim\times\xsigdim}
\safemath{\dvec}{\bmd}
\safemath{\Ddicttilde}{\widetilde{\bD}}
\safemath{\Bonb}{\bB}
\safemath{\bvec}{\bmb}
\safemath{\Bonbdim}{\ysigdim\times\ysigdim}
\safemath{\noise}{\bmn}
\safemath{\noisedim}{\ysigim}
\safemath{\err}{\bme}
\safemath{\errdim}{\ysigdim}
\safemath{\errset}{\setE}
\safemath{\nerr}{n_e}
\safemath{\delop}{\bP_\errset}
\safemath{\delopc}{\bP_{{\errset}^c}}

\safemath{\cplxi}{\imath}
\safemath{\cplxj}{\jmath}
\safemath{\dict}{\matD}
\safemath{\inputdim}{N}		% number of columns of dictionary D
\safemath{\outputdim}{M}		%number of rows of dictionary D
\safemath{\sparsity}{S}	%sparsity
\safemath{\inputdimA}{{N_a}}	%total number of elements in dictionary A
\safemath{\inputdimB}{{N_b}}	%total number of elements in dictionary B
\safemath{\elemA}{{n_a}}	%number of elements chosen from dictionary A
\safemath{\elemB}{{n_b}}	%number of elements chosen from dictionary B
\safemath{\resA}{\matR_a}	%restriction map to elements of dictionary A
\safemath{\resB}{\matR_b}	%restriction map to elements of dictionary B
\safemath{\subD}{\matS} %subdictionary
\safemath{\subA}{\matS_a} %subdictionary part of A
\safemath{\subB}{\matS_b} %subdictionary part of B
\safemath{\dicta}{\matA} 	% first subdictionary
\safemath{\dictb}{\matB} 	% second subdictionary
\safemath{\hollowS}{H}
\safemath{\hollowA}{H_a}
\safemath{\hollowB}{H_b}
\safemath{\cross}{Z}
\safemath{\coh}{\mu_d}			% coherence dictionary
\safemath{\coha}{\mu_a}			% coherence first subdictionary
\safemath{\cohb}{\mu_b}			% coherence second subdictionary
\safemath{\mubs}{\nu}	%block sub-coherence
\safemath{\cohm}{\mu_m} %mutual coherence
\safemath{\dictset}{\setD}	% set of dictionaries
\safemath{\dictsetp}{\dictset(\coh,\coha,\cohb)}	% set of dictionaries parametrized
\safemath{\dictsetgen}{\dictset_\text{gen}}
\safemath{\dictsetgenp}{\dictsetgen(\coh)}
\safemath{\dictsetonb}{\dictset_\text{onb}}
\safemath{\dictsetonbp}{\dictsetonb(\coh)}

\safemath{\leftside}{U}
\safemath{\rightsideA}{R_a}
\safemath{\rightsideB}{R_b}

\safemath{\indexS}{\setI_S} %set of indices participating in sub-dictionary S

\safemath{\na}{n_a}			% cardinality of set of linearly independent columns of first dictionary
\safemath{\nb}{n_b}			% cardinality of set of linearly independent columns of second dictionary
\safemath{\coeffa}{p_i}	%coefficients for columns of A
\safemath{\coeffb}{q_j}	%coefficients for columns of B
\safemath{\seta}{\setP}		% set of linearly independent columns of A
\safemath{\setb}{\setQ}     % set of linearly independent columns of B
\safemath{\setw}{\setW}	%set of n largest elements of w
\safemath{\setz}{\setZ}	%set of L-n largest elements of z
\safemath{\cola}{\veca}		% generic element of the dictionary A
\safemath{\colb}{\vecb}		% generic element of the dictionary B
\safemath{\cold}{\vecd}		% generic element of the dictionary D
\safemath{\inputvec}{\vecx} 	%coefficient vector (input)
\safemath{\error}{\vece}	%error vector
\safemath{\noiseout}{\vecz} 	%noisy output vector
\safemath{\inputvecel}{x}
\safemath{\inputveca}{\vecx_a}
\safemath{\inputvecb}{\vecx_b}
\safemath{\outputvec}{\vecy}	%output of Dictionary
\safemath{\lambdamin}{\lambda_{\mathrm{min}}}
\safemath{\elltwo}{\ell_2}
\safemath{\ellone}{\ell_1}
\safemath{\ellzero}{\ell_0}
\safemath{\ellinf}{\ell_\infty}
\safemath{\ellinftilde}{\ell_{\widetilde\infty}}
\safemath{\licard}{Z(\coh,\coha,\cohb)}
\safemath{\xsol}{\hat{x}}
\safemath{\xbord}{x_b}		%Solution at the border
\safemath{\xstat}{x_s}		%Solution stationary in l0 prob
\safemath{\xstatLone}{\tilde{x}_s}
\safemath{\order}{\mathcal{O}} %order notation (big O)
\safemath{\scales}{\Theta} %scales as
\safemath{\ones}{\mathbf{1}} %all ones matrix
\safemath{\zeroes}{\mathbf{0}} %all zeroes matrix
\safemath{\thlone}{\kappa(\coh,\cohb)} %treshold l1 problem
\safemath{\constoneA}{\delta} %constant in l1 theorem to save space
\safemath{\constoneB}{\epsilon} %constant in l1 theorem to save space
\safemath{\nlarge}{L}				   %num large elements
\safemath{\sumlarge}{S_\nlarge}
\safemath{\maxlarger}{P_\nlarge}	   % maximum in Gribonval and Nielsen
\safemath{\Pzero}{\textrm{P0}}	
\safemath{\Pone}{\textrm{P1}}
\safemath{\vecfir}{\vecw}			 % \vecv element of the kernel of the dictionary, \vecv=[\vecfir \vecsec]
\safemath{\vecsec}{\vecz}
\safemath{\elvecfir}{w}              % element of vecfir
\safemath{\elvecsec}{z}				 % element of vecsec
\safemath{\nlargefir}{n}
\safemath{\normout}{\gamma}
\safemath{\auxfun}{h}
\safemath{\supp}{\textrm{supp}}%support

\safemath{\indexa}{\ell}
\safemath{\indexb}{r}
\safemath{\indexc}{i}
\safemath{\indexd}{j}

\safemath{\project}{P}%projector

\usepackage{framed}

\IEEEoverridecommandlockouts
\allowdisplaybreaks 

\begin{document}

\title{Spatial and Temporal Generalization \\ of CSI-based Neural Positioning}\author{Till-Yannic M\"uller, Frederik Zumegen, Reinhard Wiesmayr, and  Christoph Studer\\[0.3cm]
\em ETH Z\"urich, Switzerland; e-mail: fzumegen@ethz.ch\thanks{This work was supported in part by the Swiss National Science Foundation (SNSF) grant 200021\_207314 and by CHIST-ERA grant for the project CHASER (CHIST-ERA-22-WAI-01) through the SNSF grant 20CH21\_218704. We acknowledge NVIDIA for its sponsorship of this research.}

}

\maketitle

\begin{abstract}

Channel state information~(CSI)-based neural positioning learns a mapping from CSI measurements to user equipment~(UE) positions using neural networks.
However, most existing performance evaluations utilize randomly partitioned train/test CSI-dataset splits, which fail to reflect the generalization requirements of practical deployments and present optimistic results.
In this paper, we study the spatial and temporal generalization of neural positioning with standard-compliant Wi-Fi and 5G~NR systems for three real-world CSI datasets acquired in indoor and outdoor environments.
We assess generalization with two different architectures, a conventional multilayer perceptron~(MLP) and a novel transformer architecture, to unseen spatial regions, unseen UE trajectories, and CSI measurement campaigns separated by one week.
Our experiments show that both architectures generalize well in space and time, and the proposed transformer consistently outperforms the MLP in positioning accuracy while requiring fewer model parameters.
\end{abstract}

\section{Introduction}
Channel state information (CSI)-based user equipment~(UE) positioning is believed to be a key technology that leverages integrated sensing and communication~(ISAC) in next-generation wireless networks~\cite{6g_localization_sensing,liu2022ISAC6G}.
While global navigation satellite systems are unreliable indoors and in dense urban environments~\cite{Granados2012GNSSIndoor} and their position estimates are unavailable to network operators, off-device positioning from measured receive signals provides a viable alternative.
In particular, neural network (NN)-based approaches have emerged as a powerful tool for off-device positioning, in which a mapping from CSI measurements to UE position is learned in a data-driven manner~\cite{Arnold2019CSIFingerprinting,Vieira2017Fingerprinting,Zhang2023FingerprintingCCN,Savic2015Figerprinting,mueller2025neuralpositioningexternalreference,Gönültaş2021probabilityFusion,gönültas2021featuresForCSIpositioning,huang2024attackDefendOffDevicePos,wiesmayr2025nvidia5gtestbed}.
Such \emph{neural positioning} methods typically rely on fully-connected multilayer perceptrons (MLPs)~\cite{Gönültaş2021probabilityFusion,gönültas2021featuresForCSIpositioning,huang2024attackDefendOffDevicePos,wiesmayr2025nvidia5gtestbed}; only recently, transformer-based architectures have been proposed for channel charting~\cite{pirkl2026resilientCCTransf,zhang2026TransfCC,park2026lightweight}.

A common shortcoming of existing work on neural positioning is the use of randomly partitioned train/test dataset splits~\cite{Arnold2019CSIFingerprinting,Vieira2017Fingerprinting,Zhang2023FingerprintingCCN,Savic2015Figerprinting,mueller2025neuralpositioningexternalreference,Gönültaş2021probabilityFusion,gönültas2021featuresForCSIpositioning,huang2024attackDefendOffDevicePos}, which does not reflect a practical use-case: in practice, a positioning system must generalize to unseen locations and temporal variations not present during training.
Reference \cite{sobehy2021generalization} is an exception and studies the generalization of an MLP and $K$-nearest-neighbor-based methods to spatially unseen regions and trajectories. Their analysis, however, is confined to a single small-scale indoor dataset covering a 4\,m$\times$2\,m table and does not consider modern neural positioning architectures, diverse wireless standards, or temporally separated measurement campaigns.

\subsection{Contributions}

We systematically study the generalization capability of neural positioning.
Our main contributions are as follows.
We investigate the efficacy of neural positioning to unseen spatial regions, unseen UE trajectories, and measurement campaigns separated by one week.
We consider two neural network architectures: a fully-connected MLP representing the state of the art and a novel transformer architecture.
We evaluate both architectures on three real-world datasets spanning indoor and outdoor environments, line-of-sight and non-line-of-sight conditions, varying measurement area sizes, and different wireless standards (Wi-Fi and 5G~NR).
Our experiments demonstrate that (i) neural positioning generalizes in both space and time and (ii) the proposed transformer architecture outperforms the MLP in terms of generalization across all datasets and with fewer model parameters.

\section{Neural Positioning}
We start by outlining the basics of neural positioning, followed by describing how (i) a fully-connected MLP and (ii) a transformer network can be utilized for this task. To arrive at a fair comparison between both network architectures, we extensively optimized their respective architectures (e.g., number of layers, activations, etc.) and training procedures.

\begin{figure*}[tp]
\centering
\includegraphics[width=0.99\textwidth]{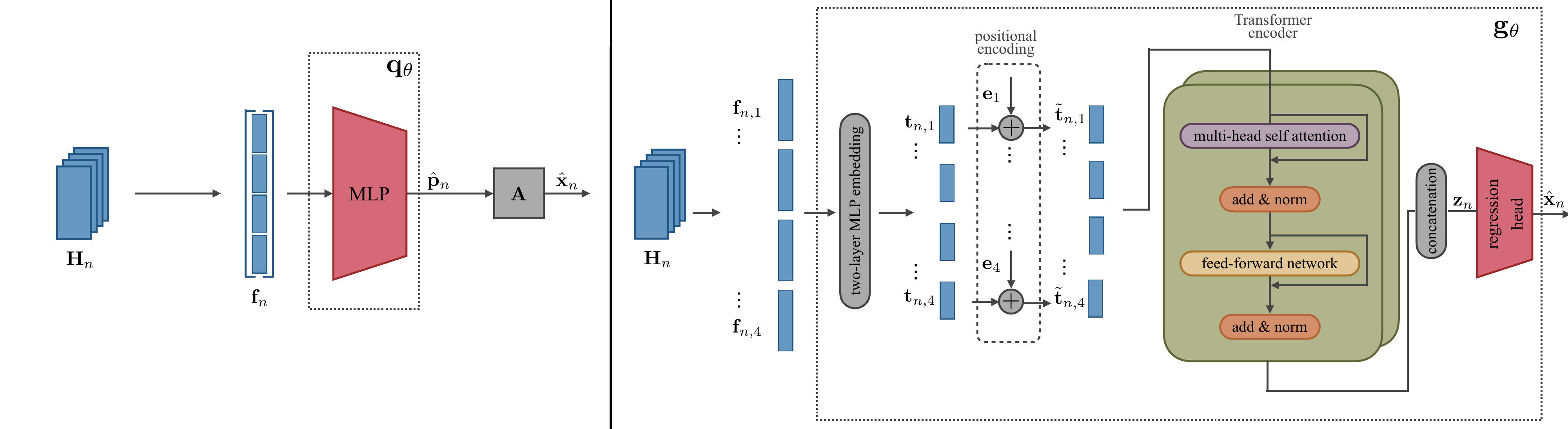}
\caption{Neural positioning architectures: fully-connected MLP trained with probability maps and BCE loss (left); transformer trained with MSE loss (right).}
\label{fig:architectures}
\end{figure*}

\subsection{Basics}

We consider an orthogonal frequency-division multiplexing~(OFDM) wireless \emph{uplink} system operating over $W$ subcarriers with a single-antenna UE transmitting to $B$ access points~(APs), each equipped with $A$ antennas.
The corresponding CSI tensor at sample instant~$n\in\{1,\ldots,N\}$ is denoted $\bH_n \in \complexset^{B \times A \times W}$. Here, $N$ denotes the total number of measured CSI samples.
From each CSI estimate~$\bH_n$, we extract the $n$th feature vector~$\bmf_n \in \reals^F$ of dimension $F=BAW$ by taking the element-wise absolute values of $\bH_n$ and normalizing its vectorized version so that $\|\bmf_n\| = 1$.

We train a positioning function $\bmg_{\bmtheta} : \reals^{F} \rightarrow \reals^{D}$ with parameters~$\bmtheta$ that maps the CSI feature vector~$\bmf_n$ to a position estimate $\hat{\bmx}_n = \bmg_{\bmtheta}(\bmf_n)$ of the corresponding ground-truth position $\bmx_n \in \reals^D$ in $D$-dimensional space (we focus on $D=2$ in this paper).
The positioning function is trained by minimizing a loss function~$\ell$ over the training dataset $\setD\subseteq \{1,\ldots,N\}$ with CSI and ground-truth-location datapoint tuples $\{(\bmf_n, \bmx_n)\}_{n\in\setD}$.

\subsubsection{MSE-based Loss}
A common choice for the loss function is the mean squared error~(MSE) loss defined as
\begin{align}
  \ell_{\text{MSE}}(\bmtheta)
  = \frac{1}{|\setD|} \sum_{n\in\setD}
  \left\| \bmx_n -
    \bmg_{\bmtheta}(\bmf_n) \right\|^2,
  \label{eq:mse}
\end{align}
which directly penalizes the square distance between the predicted position $\hat{\bmx}_n$ and the ground-truth UE position $\bmx_n$.

\subsubsection{Probability-Map-based Loss}
Another choice for the loss function put forward in~\cite{Gönültaş2021probabilityFusion} represents user positions as a probability map $\bmp_n \in [0,1]^{K}$, which is a probability mass vector at time instant~$n$.
Here, the physical space is discretized  using $K$ fixed grid-point locations $\{\bma_k\}_{k=1}^{K}$, and the UE position is given by $\bmx_n=\bA\bmp_n$ with $\bA = [\bma_1, \ldots, \bma_{K}]$.
The positioning function $\hat{\bmp}_n = \bmq_{\bmtheta}(\bmf_n)$, where $\bmq_{\bmtheta} : \reals^{F} \rightarrow \reals^{K}$ maps the CSI feature $\bmf_n \in \reals^F$ to the estimated probability map~$\hat{\bmp}_n$, is trained via a binary cross-entropy~(BCE) loss between a ground-truth probability map~$\bmp_n$ and the estimated probability map~$\hat{\bmp}_n$; see~\cite{Gönültaş2021probabilityFusion} for the details.

\subsubsection{Neural Positioning}
For both approaches, position estimates are obtained either directly from $\bmg_{\bmtheta}$ or from $\bmq_{\bmtheta}$ followed by $\hat{\bmx}_n=\bA\hat{\bmp}_n$ using CSI features from a test dataset $\setT$.

\subsection{Architecture 1: Fully-Connected MLP}
\label{sec:MLP}
The fully-connected MLP implements $\bmq_{\bmtheta}$ using the probability-map approach from~\cite{Gönültaş2021probabilityFusion}, which achieves state-of-the-art positioning performance; the architecture is illustrated on the left of \fref{fig:architectures}.
The MLP consists of four hidden layers and one output layer with dimensions $\{F,\,512,\,512,\,512,\,512,\,K\}$, where $K=484$ assuming a rectangular $22\times22$ spatial grid as in~\cite{Gönültaş2021probabilityFusion}.
We use batch normalization after the first layer; all hidden layers use ReLU activations; the output layer applies softmax, so that the entries in~$\hat{\bmp}$ sum to one.
For training, we minimize the BCE loss between $\hat{\bmp}_n$ and $\bmp_n$ (which are computed from the reference positions $\bmx_n$, $n\in\setD$), and, for testing, we obtain the final position estimate as $\hat{\bmx}_n = \bA\hat{\bmp}_n \in \reals^D,\,n\in\setT$.

\subsection{Architecture 2: Transformer}
\label{sec:transformer}
The proposed transformer implements $\bmg_{\bmtheta}$ by casting each AP as a token and processing the resulting sequence via self-attention; the architecture is illustrated on the right of~\fref{fig:architectures}.
Each AP $b$ contributes one \emph{token} $\bmt_{n,b} \in \reals^{d}$, obtained by projecting $\bmf_{n,b} \in \reals^{A W}$ through a two-layer MLP embedding with ReLU activations; a learned \emph{positional encoding} $\bme_b \in \reals^{d}$ is added to inject AP identity, giving $\tilde{\bmt}_{n,b} = \bmt_{n,b} + \bme_b$.
The token sequence $\{\tilde{\bmt}_{n,b}\}_{b=1}^{B} \in \reals^{B \times d}$ is passed through two transformer encoder layers, each consisting of multi-head self-attention with four heads followed by a feed-forward network~(FFN) of hidden dimension $d_{\mathrm{ff}}=512$; the encoder hyperparameters are summarized in \fref{tbl:hyperparameters}.
The outputs of all $B$ encoder tokens are concatenated\footnote{We note that concatenating all encoder token outputs consistently outperformed mean pooling and CLS-token extraction in our experiments.} to form $\bmz_n \in \reals^{Bd}$, which is passed through a four-layer regression head with ReLU activations and a final sigmoid activation, producing normalized position estimates\footnote{Reference position normalization was found critical for stable transformer training; The MLP did not exhibit this sensitivity.}~$\hat{\bmx}_n \in [0,1]^2$ in the normalized region $[0,1]^2$.

For transformer training, we minimize the MSE loss between~$\hat{\bmx}_n$ and ground-truth positions $\bmx_n$ normalized to the region $[0,1]^2$; the final position estimate is recovered by inverting the normalization step.
We note that a probability-map-based output head with BCE-based training resulted in inferior performance compared to MSE-based training.

\begin{figure*}[t]
    \centering
    \includegraphics[width=0.99\textwidth]{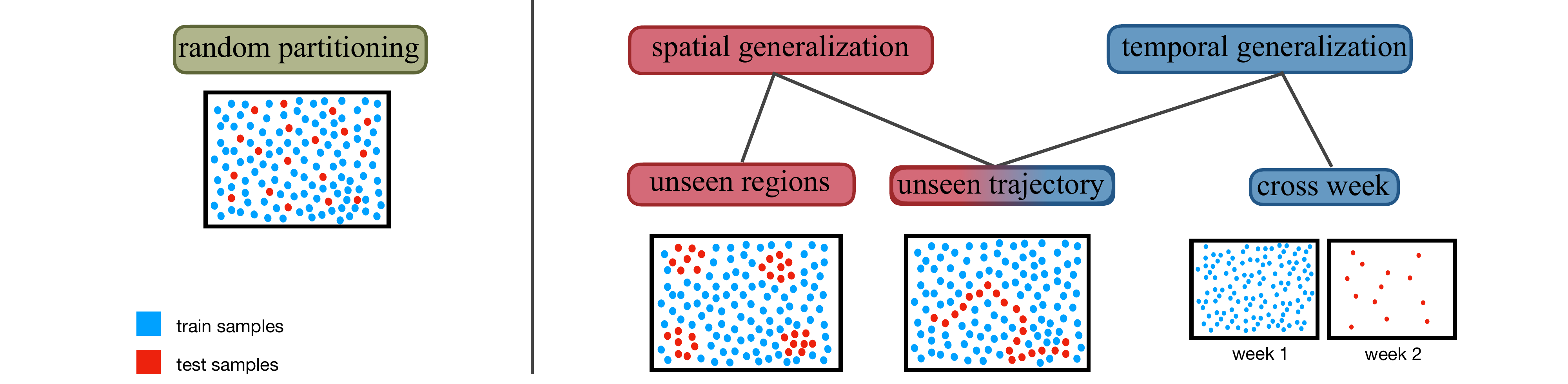}
    \caption{Illustration of the used testing scenarios for assessing positioning performance:
    conventional random partitioning baseline~(left);
    proposed testing scenarios for assessing spatial and temporal generalization of neural positioning~(right).}
    \label{fig:metric_sketch}
\end{figure*}

\begin{table}[t]
\centering
\caption{Transformer encoder hyperparameters}
\label{tbl:hyperparameters}
\renewcommand{\arraystretch}{1.15}
\begin{tabular}{@{}lc@{}}
\toprule
\textbf{Parameter} & \textbf{Value} \\
\midrule
Embedding dimension ($d$) & 128 \\
Number of encoder layers               & 2   \\
Number of attention heads              & 4   \\
FFN hidden dimension ($d_\text{ff}$)   & 512 \\
FFN activation                         & ReLU \\
Dropout rate                           & 0.1 \\
\bottomrule
\end{tabular}
\end{table}

\subsection{Training}
The training parameters are summarized in \fref{tbl:training_parameters}.
Both architectures use Kaiming uniform weight initialization~\cite{he2015}; the MLP is optimized with Adam~\cite{kingma2015adam} with training parameters summarized in \fref{tbl:training_parameters}.
The MLP employs a step-decay learning rate schedule, while the transformer uses cosine annealing with a minimum learning rate of $10^{-6}$.
The MLP is trained for $30$ epochs with a batch size of $B_{\text{batch}}=10$ minimizing the BCE loss between the predicted probability map $\hat{\bmp}_n$ and its ground-truth counterpart $\bmp_n$; the transformer is trained for $1000$ epochs\footnote{ Training the MLP beyond $30$ epochs yields no further improvement.} with a batch size of $B_{\text{batch}}=32$ by minimizing~\fref{eq:mse} on normalized coordinates.
All models are trained on an NVIDIA RTX~4070 GPU using PyTorch~\cite{paszke2019pytorch}.
As shown in \fref{tbl:training_parameters}, the transformer has fewer than half the number of parameters compared to the MLP.

\begin{table}[tp]
\centering
\caption{Summary of training parameters.}
\label{tbl:training_parameters}
\renewcommand{\arraystretch}{1.15}
\begin{tabular}{@{}lcc@{}}
\toprule
\textbf{Parameter} & \textbf{MLP} & \textbf{Transformer} \\
\midrule
Weight initialization      & Kaiming uniform               & Kaiming uniform   \\
Loss function               & BCE                           & MSE               \\
Optimizer                   & Adam                          & AdamW             \\
Weight decay                & n.a.                          & $10^{-4}$  \\
Learning rate               & $10^{-4}$              & $10^{-4}$  \\
LR scheduler                & step decay                    & cosine annealing  \\
Minimum learning rate       & n.a.                           & $10^{-6}$  \\
Epochs                      & $30$                            & $1000$              \\
Batch size                  & $10$                            & $32$                \\
Trainable param. (5G)   & $3\,274\,212$                 & $1\,098\,402$     \\
Trainable param. (Wi-Fi)& $1\,888\,740$                 & $752\,290$        \\
\bottomrule
\end{tabular}
\end{table}

\section{Metrics and Generalization Scenarios}

We now introduce the error metrics used for evaluation and the considered baseline and generalization scenarios.

\subsection{Positioning Error Metrics}

We evaluate positioning performance using three metrics, which are all based on the absolute error~(AE) between estimated position and ground-truth position: $e_n = \|\bmx_n - \hat{\bmx}_n\|$.
All metrics are reported in meters; an overview of all testing scenarios is provided in \fref{fig:metric_sketch}.
The mean absolute error~(MAE) averages the AE over all samples from the test dataset $\setT$:
\begin{align}
  \textit{MAE} = \frac{1}{|\setT|} \sum_{n\in\setT}  e_n.
  \label{eq:mae}
\end{align}
The median error is the $50$th percentile of $\{e_n\}_{n\in\setT}$, providing a measure of typical positioning performance robust to outliers.
The $95$th-percentile error is the value $\bar{e}_{95}$ such that $95\%$ of the AEs in the test dataset $\setT$ satisfy $e_n \leq \bar{e}_{95}$.

As a reference, we also include a single-best-guess baseline with respect to the geometric median of the training positions, i.e., the point $\bmx^\star\in\reals^2$ minimizing the sum of distances to all training samples as follows:
\begin{align}
  \bmx^\star = \arg\min_{\bmc \in \mathbb{R}^2}
  \sum_{n\in\setD} \|\bmx_n - \bmc\|.
  \label{eq:geomed}
\end{align}
Since~\fref{eq:geomed} admits no closed-form solution, we numerically compute $\bmx^\star$ via Weiszfeld's algorithm~\cite{weiszfeld1937singleguess}.
By construction,~$\bmx^\star$ minimizes the MAE over all single-best-guess predictors.

\subsection{Baseline Scenario: Random Partitioning}

As a baseline scenario, we randomly partition the full dataset into $80\%$ for training and retain the remaining $20\%$ for testing; this split serves as the reference against which all subsequent generalization scenarios are compared. Note that this is the de-facto standard scenario used to evaluate neural positioning methods~\cite{Arnold2019CSIFingerprinting,Vieira2017Fingerprinting,Zhang2023FingerprintingCCN,Savic2015Figerprinting,mueller2025neuralpositioningexternalreference,Gönültaş2021probabilityFusion,gönültas2021featuresForCSIpositioning,huang2024attackDefendOffDevicePos}.
\fref{fig:metric_sketch} (left) illustrates the resulting training and test sample distributions.

\subsection{Generalization Scenarios in Space and Time}

\subsubsection{Unseen Regions}
In order to assess spatial generalization, we withhold multiple circular regions of radius $r = 0.44$\,m (which is $5\times$ the wavelength of the 5G~NR system) entirely from training and use them exclusively for testing.
Thereby, the neural network never observes CSI from the UE located within these test regions during training.
\fref{fig:metric_sketch} (middle-left) illustrates the corresponding training and testing sample distributions.
\subsubsection{Unseen Trajectory}
In order to assess temporal generalization, we withhold a contiguous UE trajectory entirely from training and use it exclusively for testing.
Thereby, the neural network never observes CSI along this UE trajectory during training.
\fref{fig:metric_sketch} (middle-right) illustrates the corresponding training and testing sample distributions.
For each measured dataset (described in \fref{sec:measuredCSIdatasets}), the unseen trajectory comprises 1\% of the total dataset (roughly 1\,min CSI acquisition) and is extracted from the center of each measurement sequence.
The trajectory lengths are 21\,m (5G Hallway), 41\,m (5G Outdoor), 13\,m (Wi-Fi Office Week 1), and 8\,m (Wi-Fi Office Week 2).

\subsubsection{Cross-Week Consistency}
In order to assess temporal generalization over larger time spans, we conduct two measurement campaigns in the same office environment one week apart, and we evaluate three configurations: (i)~training on data from Week~$1$ and testing on Week~$2$, (ii)~training on data from Week~$2$ and testing on Week~$1$, and (iii)~training and testing on randomly partitioned data from both weeks.
\fref{fig:metric_sketch} (right) illustrates the corresponding training and testing sample~distributions.

\section{Measured CSI Datasets}
\label{sec:measuredCSIdatasets}
In order to assess the generalization capability of the two proposed neural positioning architectures, we consider three CSI datasets from real-world over-the-air measurements.
The datasets cover Wi-Fi and 5G~NR deployments across three indoor and outdoor environments.
The Wi-Fi and 5G testbeds used to extract CSI as well as ground-truth position information are described in detail in~\cite{zumegen2024wifiCSI} and~\cite{wiesmayr2025nvidia5gtestbed}, respectively.
The key properties of each dataset are summarized in \fref{tbl:datasets}.
For the two 5G datasets, the full $3\,276$ subcarriers are first low-pass filtered and then downsampled by a factor of $12$ prior to training, yielding $273$ active subcarriers; we found that using the full number of subcarriers yields no performance gains.

\begin{figure}[t]
  \centering
  \subfigure[]{
    \includegraphics[height=3.0cm]{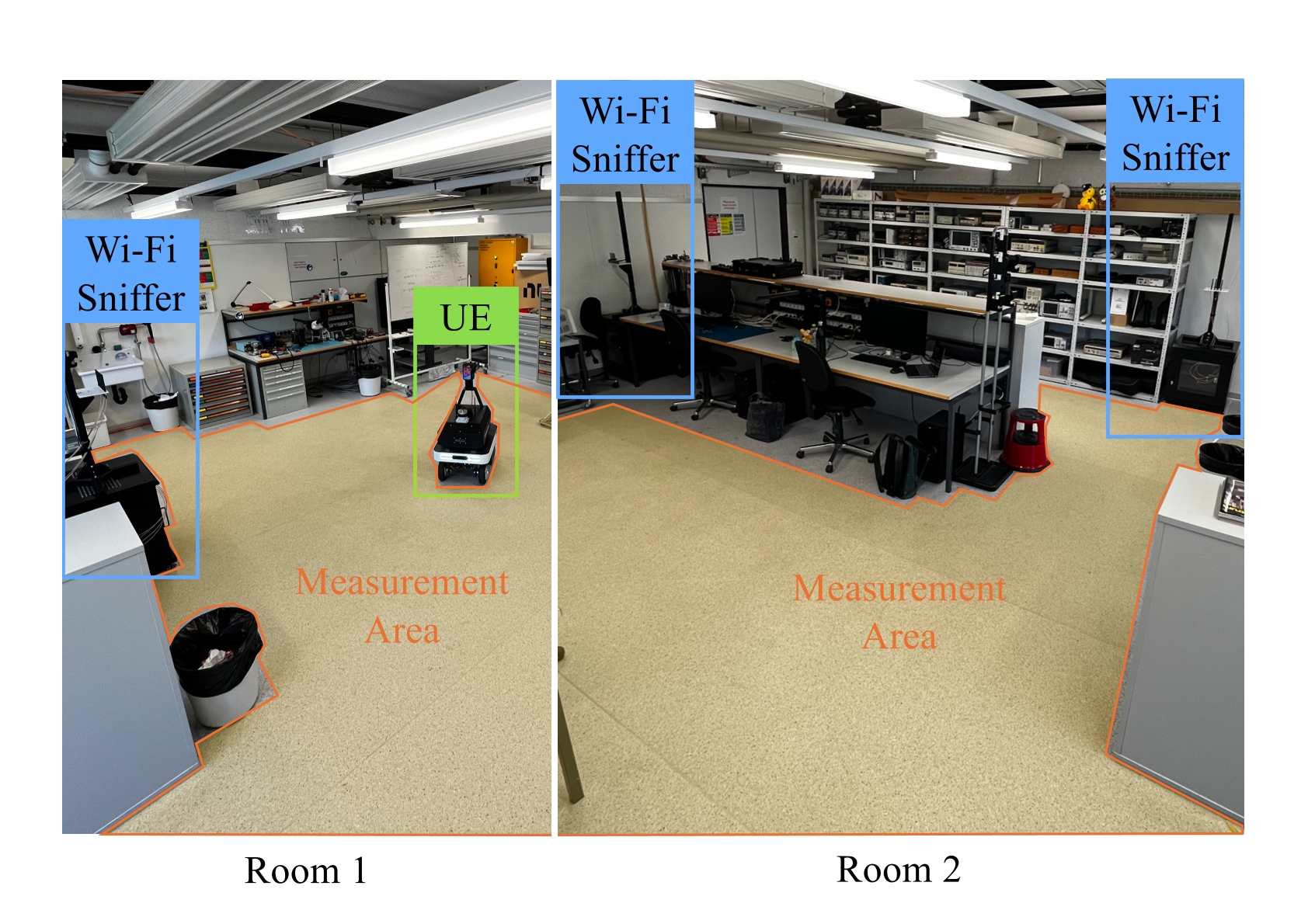}\label{fig:wifi_office_photo}%
  }\hfill
  \subfigure[]{
    \includegraphics[height=3.0cm]{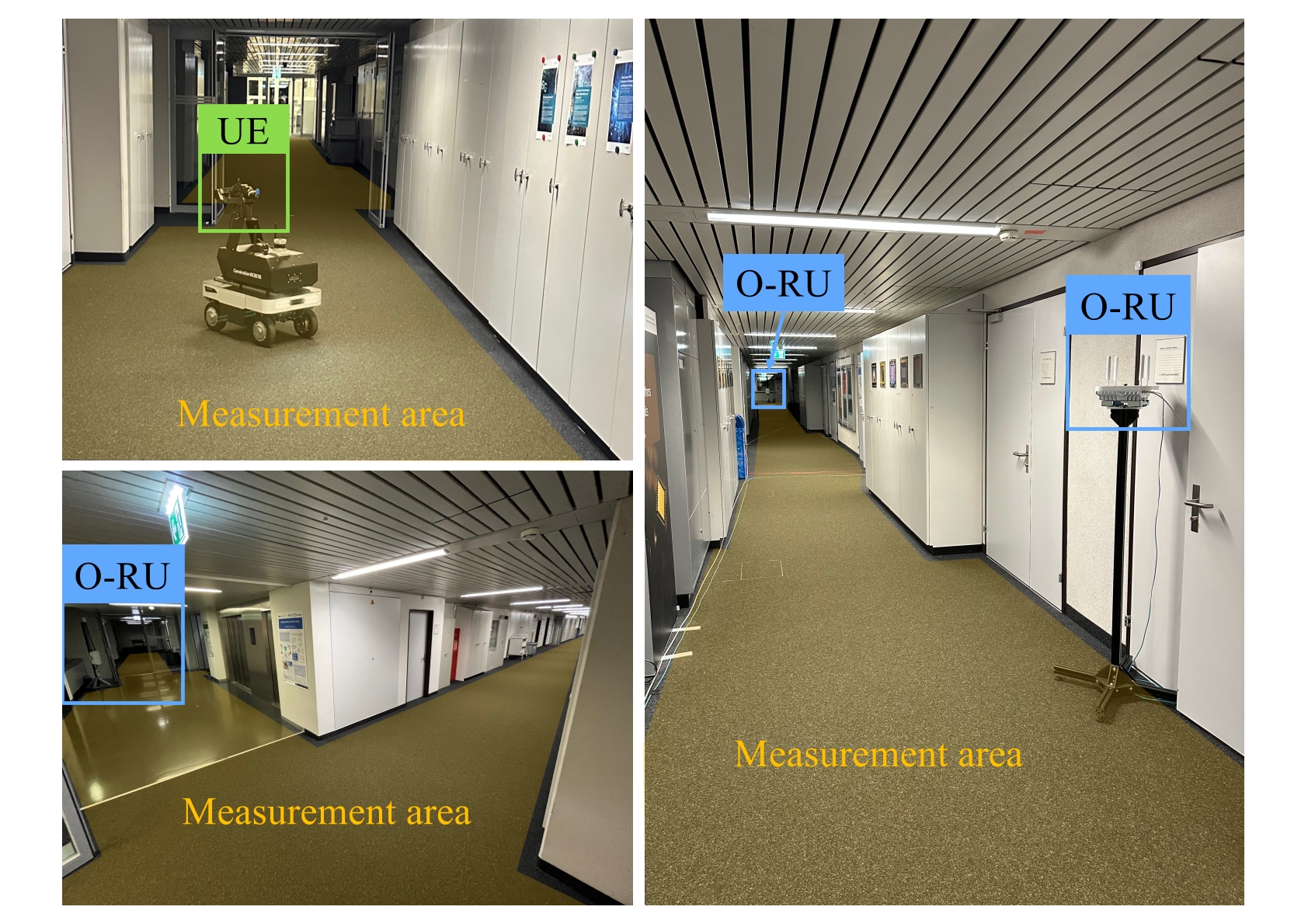}\label{fig:jfloor_photo}%
  }
  \subfigure[]{
    \includegraphics[height=3.0cm]{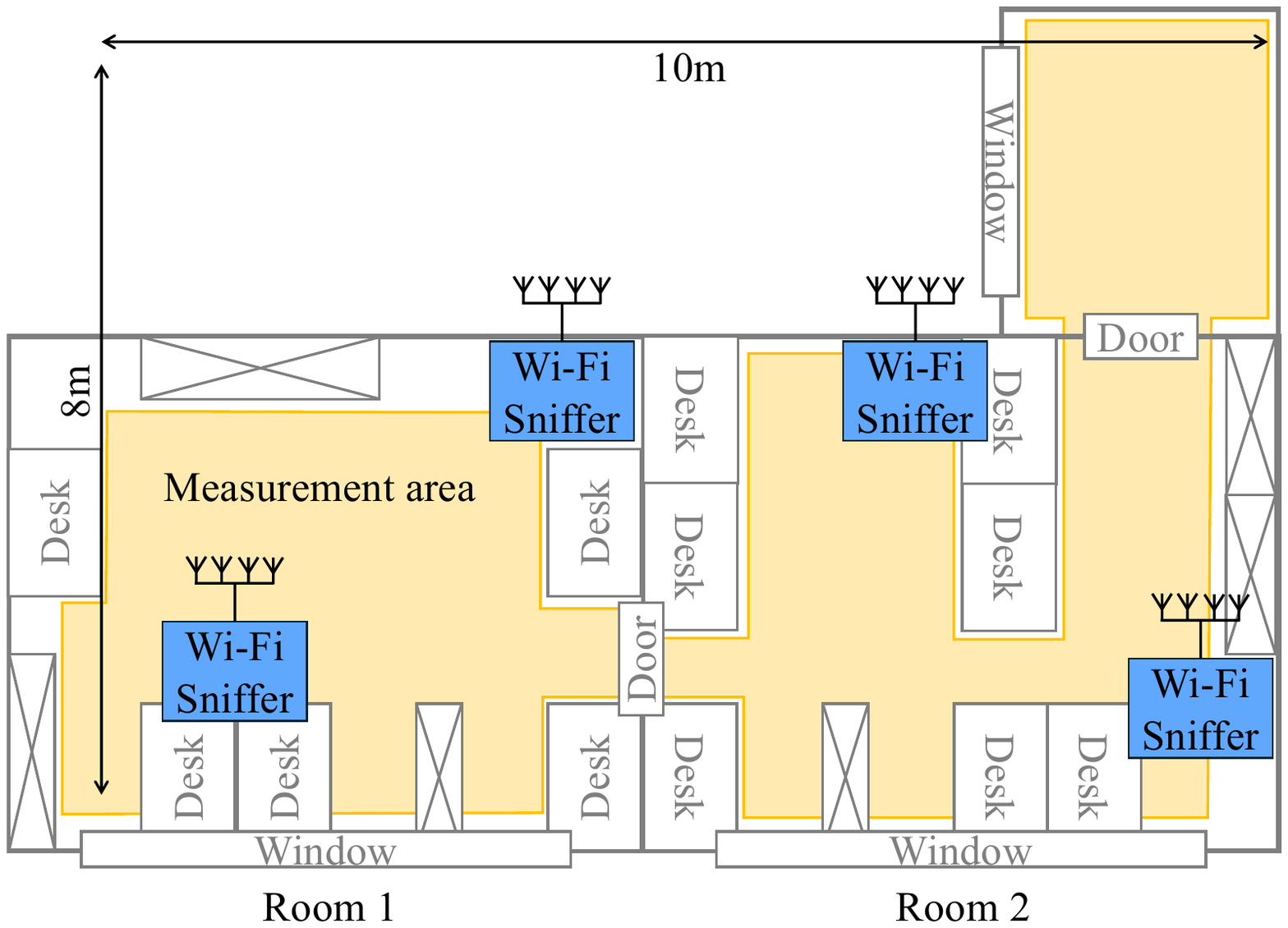}\label{fig:wifi_office_floorplan}%
  }\hfill
  \subfigure[]{
    \includegraphics[height=3.0cm]{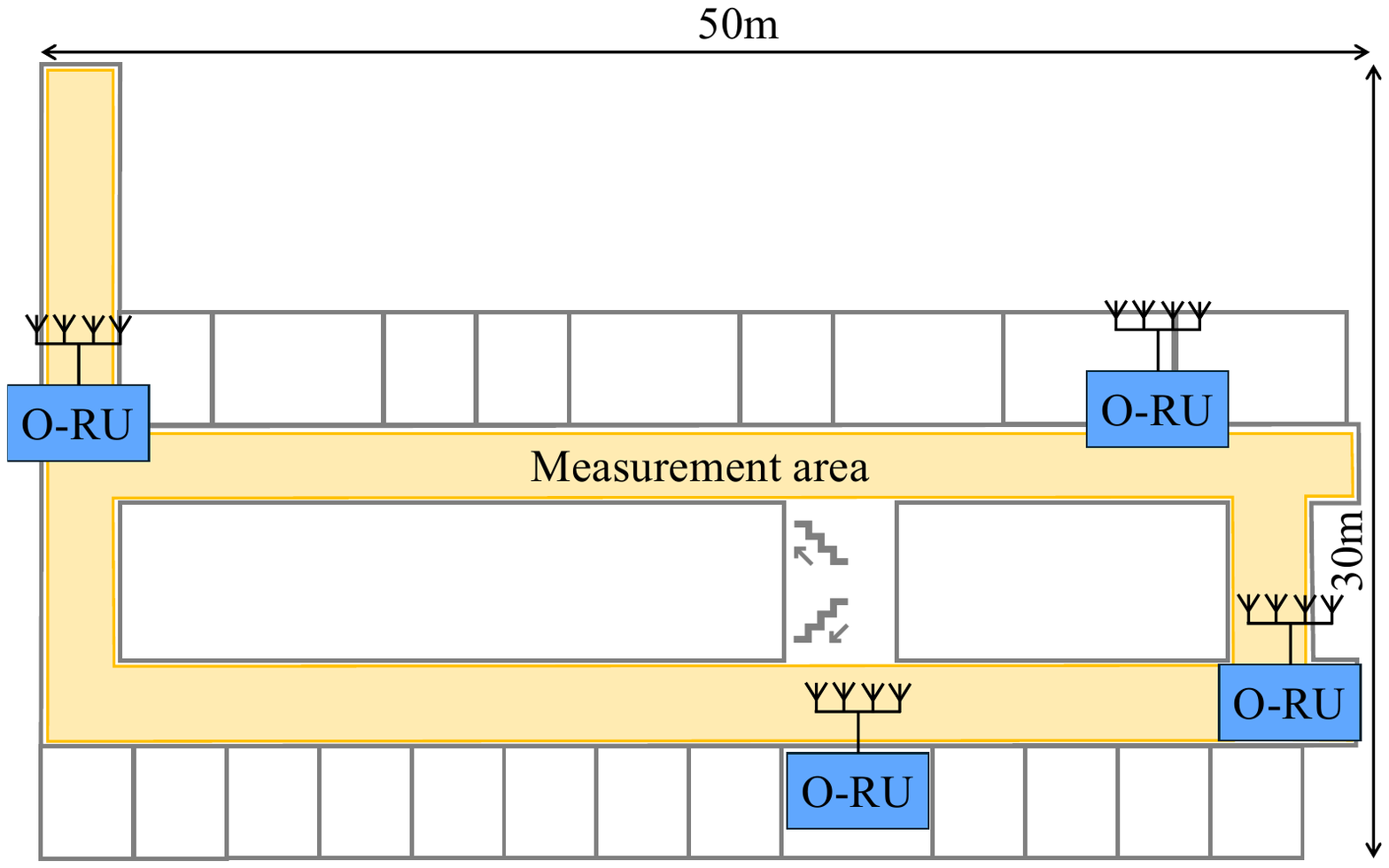}\label{fig:jfloor_floorplan}%
  }
  \caption{Measurement areas (top) and floorplans (bottom) for the new Wi-Fi Office dataset~(left) and the new 5G Hallway dataset~(right).}
  \label{fig:environments}
\end{figure}

\subsection{New CAEZ-WIFI-OFFICE-2WEEKS Dataset}

The ``Wi-Fi Office'' dataset was acquired with our Wi-Fi testbed in an indoor office space in a measurement area of $13.7$\,m $\times$ $10.6$\,m with $4$ APs and $4$ antennas per AP.
The dataset consists of two measurement campaigns conducted in the same environment with a separation of one week, each of which was recorded for slightly more than one hour at a CSI sampling rate of $17$\,Hz.
We collected CSI at all APs from the Wi-Fi uplink channel and used all $104$ active subcarriers.
Note that in the week between the two measurement campaigns, APs and other objects were slightly displaced due to regular laboratory work.
During both measurement campaigns, multiple humans were present (and moving) in the close vicinity of the UE and APs.
\fref{fig:environments} depicts the floorplan and measurement area.

\subsection{New CAEZ-5G-HALLWAY Dataset}

The ``5G Hallway'' dataset was acquired with our 5G NR testbed indoors on a large floor spanning $52.8$\,m $\times$ $29.4$\,m, with $4$ 5G Open RAN Radio Units (O-RUs) and $4$ antennas per O-RU.
The dataset was recorded over more than one hour at a CSI sampling rate of $51$\,Hz.
As for the ``Wi-Fi Office'' dataset, multiple humans were present (and moving) during the measurement campaign.
\fref{fig:environments} depicts the floorplan and measurement area.

\subsection{Existing CAEZ-5G-OUTDOOR Dataset}

The ``5G Outdoor'' dataset is an existing dataset that was acquired with our 5G NR testbed outdoors in a courtyard in a measurement area of size $9.9$\,m $\times$ $10.1$\,m, with $4$ O-RUs and $4$ antennas per O-RU.
The dataset was recorded over more than one hour at a CSI sampling rate of $51$\,Hz.
For a detailed description of this dataset, we refer the reader to~\cite{wiesmayr2025nvidia5gtestbed}.

\begin{table}[tp]
\centering
\caption{Overview of CSI Datasets.}
\label{tbl:datasets}
\renewcommand{\arraystretch}{1.15}
\resizebox{\columnwidth}{!}{%
\begin{tabular}{@{}lccc@{}}
\toprule
 & \textbf{Wi-Fi} \textbf{Office} & \textbf{5G} \textbf{Hallway} & \textbf{5G} \textbf{Outdoor} \\
\midrule
Wireless standard & IEEE 802.11a  & 5G NR         & 5G NR         \\
Carrier freq.     & 5.23\,GHz     & 3.45\,GHz     & 3.45\,GHz     \\
Bandwidth         & 40\,MHz       & 100\,MHz      & 100\,MHz      \\
Samples $N$           & 113\,716\,/ 96\,170\textsuperscript{$a$} & 288\,180      & 301\,172      \\
Duration          & 1\,h\,51\,min\,/ 1\,h\,36\,min\textsuperscript{$a$} & 1\,h\,35\,min   & 1\,h\,38\,min   \\
Sample rate       & 17\,Hz        & 51\,Hz        & 51\,Hz        \\
APs $B$              & 4             & 4             & 4             \\
Antennas $A$      & 4             & 4             & 4             \\
Active subcarriers           & 104           & 3\,276        & 3\,276        \\
Downsampling factor        & n.a.             & 12$\times$    & 12$\times$   \\
Used subcarriers $W$         & 104           & 273           & 273           \\
Covered area              & $13.7{\times} 10.6$\,m$^2$ & $52.8{\times} 29.4$\,m$^2$ & $9.9{\times}10.1$\,m$^2$ \\
\bottomrule
\multicolumn{4}{l}{\textsuperscript{$a$}Week\,1 / Week\,2.}
\end{tabular}}
\end{table}

\begin{figure*}[tp]
\centering
\subfigure[Wi-Fi Office Week 1]{%
    \includegraphics[width=0.48\textwidth]{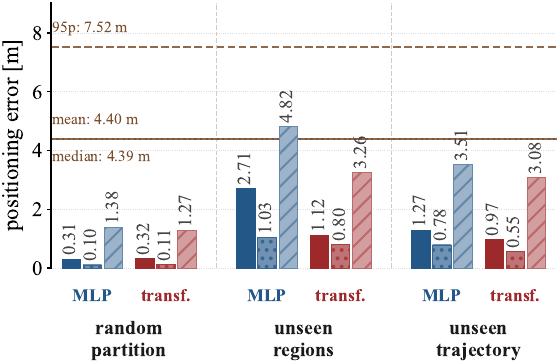}%
    \label{fig:arch_wifi1}%
}\hfill
\subfigure[Wi-Fi Office Week 2]{%
    \includegraphics[width=0.48\textwidth]{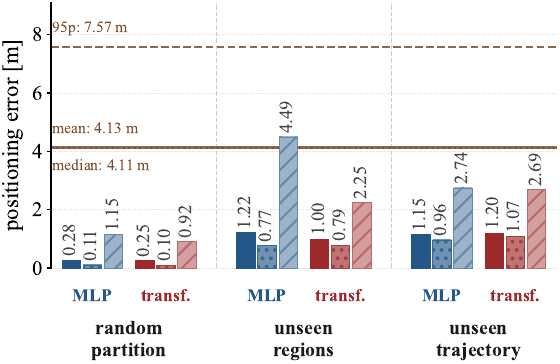}%
    \label{fig:arch_wifi2}%
}\\[4pt]
\subfigure[5G Hallway]{%
    \includegraphics[width=0.48\textwidth]{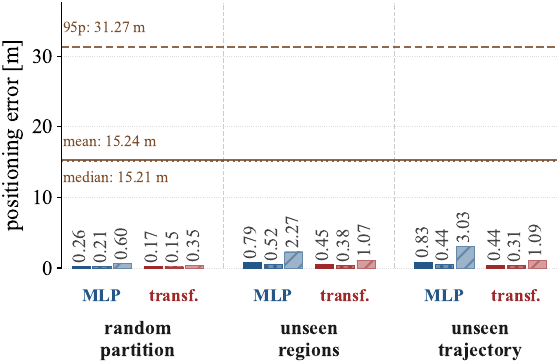}%
    \label{fig:arch_jfloor}%
}\hfill
\subfigure[5G Outdoor]{%
    \includegraphics[width=0.48\textwidth]{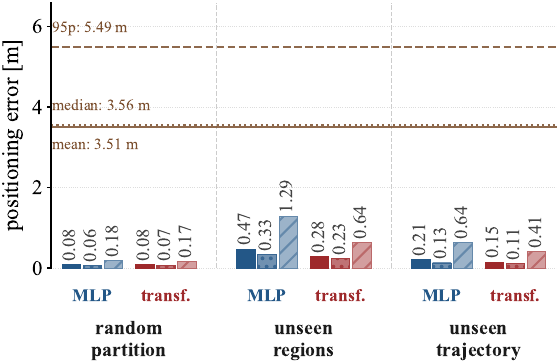}%
    \label{fig:arch_caez}%
}\\[4pt]
\includegraphics[width=0.6\textwidth]{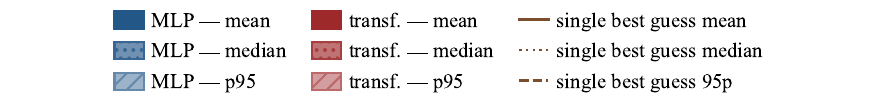}
\caption{Positioning error for the MLP (blue) and transf.\ (red) architectures across four datasets and three testing scenarios: random partitioning, unseen regions, and unseen trajectory; mean, median, and 95th-percentile errors are reported. Both architectures remain consistently below the single-best-guess baseline (brown), and the transf.\ outperforms the MLP across all datasets despite having fewer parameters.}
\label{fig:arch_comparison}
\end{figure*}

\begin{figure*}[tp]
    \centering
    \includegraphics[width=\textwidth]{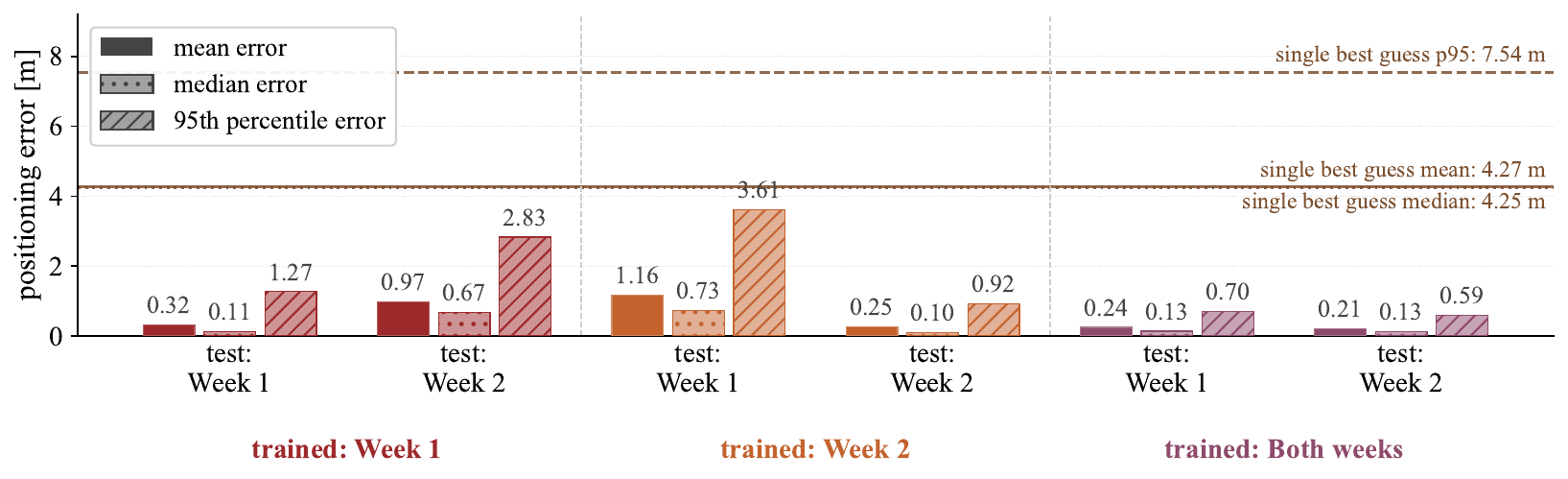}
    \caption{Positioning error for the transformer architecture grouped by training set on the Wi-Fi Office 
    Dataset; mean, median, and 95th-percentile errors are shown for 
    same-week testing, cross-week testing, and joint training and testing 
    on data from both weeks. 
    Single-week training suffers a significant performance degradation when 
    evaluated on the held-out week; joint training on both weeks 
    yields the lowest errors across all metrics.}
    \label{fig:cross_week_positioning}
\end{figure*}

\section{Results}

We now evaluate the generalization capabilities of the fully-connected MLP and the transformer introduced in \fref{sec:MLP} and \fref{sec:transformer}, respectively, and benchmark against two baselines: the single-best-guess (predictor baseline) and random partitioning (testing scenario baseline).
\fref{fig:arch_comparison} summarizes the positioning accuracy for both proposed architectures across all datasets: the Wi-Fi Office dataset (Week~1 and Week~2, top left and right, respectively), the 5G Hallway dataset (bottom left), and the 5G Outdoor dataset (bottom right). The single-best-guess baseline is shown as brown horizontal lines, and results for random partitioning, unseen regions, and an unseen trajectory are shown in each subfigure.

\subsection{Spatial Generalization}

The mean, median, and 95th-percentile positioning errors on the unseen regions are reported in \fref{fig:arch_comparison}.
Positioning performance on unseen regions falls below that of random partitioning across all datasets. This confirms that random partitioning produces overly optimistic results and does \emph{not} reflect realistic deployment conditions.
Despite this performance gap, both considered neural positioning architectures demonstrate meaningful generalization to unseen regions, with positioning errors remaining noticeably and consistently below the single-best-guess baseline.
The transformer outperforms the fully-connected MLP on all datasets despite having fewer parameters.
Moreover, we see that the 5G datasets yield lower positioning errors than their Wi-Fi counterparts.

\subsection{Temporal Generalization}

The mean, median, and 95th-percentile positioning errors on the unseen trajectories are reported in \fref{fig:arch_comparison}.
Positioning performance on the unseen trajectories falls below that of random partitioning but remains consistently below the single-best-guess baseline across all datasets, again confirming that both architectures retain meaningful localization capability under temporal domain shift.
Again, the transformer consistently outperforms the MLP on all three datasets.

\fref{fig:cross_week_positioning} shows the cross-week positioning performance only of our proposed transformer\footnote{The MLP exhibits slightly higher positioning errors but demonstrates identical qualitative trends as the proposed transformer.} reporting results for training on the Wi-Fi Office dataset Week~1 only ({red}), Week~2 only ({orange}), and both weeks jointly ({purple}).
Single-week training leads to a significant performance degradation on the held-out week; both architectures nonetheless outperform the single-best-guess baseline, indicating that temporal domain shift reduces but does not eliminate localization capability.
Furthermore, we see that joint training on both measurement weeks consistently outperforms single-week training across all metrics and architectures, demonstrating that temporal generalization is best addressed by augmenting the training set with measurements over large time periods; such augmentation also improves robustness to slight variations in the RF environment.

\section{Conclusions}

We have investigated the generalization capabilities of neural positioning to spatially unseen regions, temporally unseen trajectories, and measurement campaigns separated by one week. Both the fully-connected MLP and the proposed transformer architecture achieve strong results across all three scenarios.
Our results have shown that the transformer outperforms the MLP in all generalization scenarios despite having fewer trainable parameters, demonstrating that the token-based, per-AP processing of CSI features yields more generalizable representations than traditional neural positioning architectures.

A key limitation of neural positioning is the requirement 
for labeled reference position measurements during training. However, our results indicate that measurements over long time spans significantly improve positioning performance.
Thus, eliminating the need for labeled references, e.g., via the recent relative positioning approach in~\cite{mueller2025neuralpositioningexternalreference}, would enable such long-term measurements and potentially improve neural positioning and its generalization capabilities. 

Upon acceptance of this paper, we will make the new Wi-Fi Office and 5G Hallway datasets, as well as the code to perform our experiments, available to the public.

\balance

\balance

\end{document}